%% file: arxiv_v0.tex
\documentclass[preprint,sort&compress,12pt]{elsarticle}

\input{preamble.macros.tex}

\input{preamble.tikz.tex}

\usepackage{tikz}
\usepackage[utf8]{inputenc}
\usepackage{pgfplots}
\pgfplotsset{compat=newest}
\usepgfplotslibrary{groupplots}
\usepgfplotslibrary{dateplot}
\usepackage{tikzscale}
\usetikzlibrary{arrows,automata}
\usepackage{diagbox} 

\usepackage{xcolor}
\usepackage[algo2e]{algorithm2e}
\usepackage{amssymb}
\usepackage{amsthm}
\usepackage{amsmath}
\usepackage{mathrsfs}
\usepackage{mathtools}
\usepackage{textcomp}
\usepackage{gensymb}
\usepackage{graphicx}
\usepackage{subfig}
\usepackage{float}
\usepackage{caption}
\usepackage{tabu}
\usepackage{algorithm}
\usepackage{algpseudocode}
\usepackage{multirow}
\usepackage{enumerate}
\usepackage{array}
\usepackage[flushleft]{threeparttable}
\usepackage{lineno}
\usepackage{fullpage}
\usepackage{xcolor}
\usepackage{todonotes}
\usepackage{bbm}
\usepackage{listings}
\usepackage[colorlinks=true]{hyperref}
\usepackage{url}
\usepackage{siunitx}
\usepackage{makecell}
\usepackage{listings}

\usepackage[makeroom]{cancel}

\graphicspath{ {./figs/} }
\biboptions{comma,square}
\graphicspath{ {./figs/} }

\journal{Elsevier}

\theoremstyle{definition}

\theoremstyle{remark}
\newtheorem*{remark}{Remark}

\usepackage{changes}
\definechangesauthor[name={Reviewer 1}, color = orange]{Gao}
\definechangesauthor[name={Reviewer 2}, color = red]{Zahr}
\definechangesauthor[name={Reviewer 1 \& 2}, color = olive]{All}
\definechangesauthor[name={Reviewer 3}, color = brown]{Wang}
\definechangesauthor[name={Reviewer 4}, color = green]{R4}

\newcommand{\pvar}{u} 

\begin{document}
\begin{frontmatter}

\title{Physics-informed graph neural Galerkin networks: A unified framework for solving PDE-governed forward and inverse problems}
\author[ndAME]{Han Gao}
\author[ndAME]{Matthew J. Zahr}
\author[ndAME]{Jian-Xun Wang\corref{1stcor}}

\address[ndAME]{Department of Aerospace and Mechanical Engineering, University of Notre Dame, Notre Dame, IN}

\cortext[1stcor]{Corresponding authors.\ \texttt{jwang33@nd.edu} (J.-X. Wang)}

\begin{abstract}
Despite the great promise of the physics-informed neural networks (PINNs) in solving forward and inverse problems, several technical challenges are present as roadblocks for more complex and realistic applications. First, most existing PINNs are based on point-wise formulation with fully-connected networks to learn continuous functions, which suffer from poor scalability and hard boundary enforcement. Second, the infinite search space over-complicates the non-convex optimization for network training. Third, although the convolutional neural network (CNN)-based discrete learning can significantly improve training efficiency, CNNs struggle to handle irregular geometries with unstructured meshes. To properly address these challenges, we present a novel discrete PINN framework based on graph convolutional network (GCN) and variational structure of PDE to solve forward and inverse partial differential equations (PDEs) in a unified manner. The use of a piecewise polynomial basis can reduce the dimension of search space and facilitate training and convergence. Without the need of tuning penalty parameters in classic PINNs, the proposed method can strictly impose boundary conditions and assimilate sparse data in both forward and inverse settings. The flexibility of GCNs is leveraged for irregular geometries with unstructured meshes. The effectiveness and merit of the proposed method are demonstrated over a variety of forward and inverse computational mechanics problems governed by both linear and nonlinear PDEs.

\end{abstract}

\end{frontmatter}

\section{Introduction}
Partial differential equations (PDEs) play an important role in engineering applications since most of the physics governing natural or man-made complex systems are described by PDEs. However, finding solutions to most PDEs is a challenging problem, which may involve sophisticated numerical techniques and can be time-consuming, particularly for scenarios where parameters or initial/boundary conditions are partially known. Most recently, physics-informed neural networks (PINNs)~\cite{raissi2019physics}, as a new paradigm for solving both forward and inverse PDEs, has attracted increasing attention due to its great flexibility and simplicity compared to classic numerical methods. The general idea of PINNs is to approximate the PDE solutions with deep neural networks, whose loss functions are formulated as a combination of PDE residuals and data mismatch. This unique loss formulation enables physics-informed training that leverages the information from both physics equations and sparse observation data. 


Based on how to construct differential operators of PDE residuals using neural networks, PINNs can be classified into two categories: continuous and discrete. The continuous PINNs usually employ fully-connected (FC) neural networks to approximate the continuous solution function $f(\mathbf{x}, t)$ with respect to spatiotemporal coordinates $(\mathbf{x}, t)$ in a point-wise manner, where the spatial and temporal derivative terms are computed using automatic differentiation (AD) techniques~\cite{baydin2018automatic}. The continuous PINNs are undergoing a renaissance since recent impressive contributions made by Raissi et al.~\cite{raissi2019physics} on development of the continuous FC-PINN for solving forward and inverse PDEs. Its merit and effectiveness has been demonstrated over a plethora of scientific applications in many areas~\cite{chen2020physics,lu2021deepxde,rao2021physics,chen2020deep,sahli2020physics}. For instance, in fluid applications, PINNs have been used for fast surrogate modeling of idealized vascular flow problems in a forward parametric setting without training labels~\cite{sun2020surrogate}. Moreover, PINNs have also been formulated in an inverse modeling setting to
extract unobservable information (e.g., blood flow velocity) from observable data (e.g., concentration data) in cardiovascular problems~\cite{raissi2020hidden,kissas2020machine,cai2021artificial,arzani2021uncovering}. Jin et al.~\cite{jin2021nsfnets} applied FC-PINNs to solve Navier-Stokes equations, ranging from laminar to turbulent regimes, while Mao et al.~\cite{mao2020physics} further showed their effectiveness on high-speed flow problems. Recently, NVIDIA developed a scalable implementation SimNet based on continuous PINNs and applied it to solve various multiphysics problems with massive GPU parallelization~\cite{hennigh2020nvidia}. 


Despite the enormous success and rapid developments thanks to their great flexibility, the current continuous PINNs still have some limitations. First, they suffers from high training cost since the point-wise formulation requires huge amount of AD computations on vast collocation points in a high-dimensional spatiotemporal (and parameter) domain~\cite{wang2020understanding,jagtap2020conservative}. Second, it is challenging to formulate a strict enforcement of initial/boundary conditions (IC/BCs) for continuous PINNs, which has been demonstrated to be effective in finding correct unique PDE solutions, especially when labeled data is very scarce or absent~\cite{sun2020surrogate}. Although a distance-based particular solution can be introduced to strictly impose IC/BCs on a few simple 2-D domains using either specifically designed algebraic expressions or low-capacity neural networks~\cite{sun2020surrogate,berg2018unified}, it fails to show the effectiveness on complex geometries for real-world applications. To reduce training costs and enable efficient learning, discrete PINNs that leverage convolution operations and numerical discretizations have begun to spur interests due to their better efficiency and scalability~\cite{zhu2019physics, gao2020phygeonet}. Specifically, convolutional neural networks (CNN) are often used in discrete PINN to directly learn the entire spatiotemporal solution fields end to end and all the derivative terms of the physics-informed loss are calculated based on numerical discretization instead of point-wise AD. For instance,  Zhu et al.~\cite{zhu2019physics} developed a physics-constrained convolutional aencoder-decoder to solve high-dimensional elliptic PDEs, and Geneva et al.~\cite{geneva2020modeling} further extended this framework to dynamic hyperbolic PDEs with parametric initial conditions. Zhang et al.~\cite{zhang2020physics} presented a physics-guided CNN for seismic response modeling and also explored the similar idea with a Recurrent Neural Network (RNN) for metamodeling of nonlinear structures~\cite{zhang2020physics}. Wandel et al.~\cite{wandel2021teaching} recently proposed a data-free fluid surrogate based on an autoregressive U-net in a parametric setting. In aforementioned works, the computational domains are regular and discretized by uniform grids, where PDE residuals are calculated by finite difference (FD) methods. This is because the FD-based CNNs are fundamentally rooted in structured Cartesian grids of rectangular domains. Besides FD-based PINN, finite volume (FV) discretization has also been utilized to construct the PDE-based loss function to solve steady fluid problems, which, however, is still restricted to rectangular domains due to intrinsic limitations of classic convolution operations~\cite{ranade2021discretizationnet}. To enable physics-informed CNNs to solve parametric PDEs on irregular domains with unstructured grids, Gao et al.~\cite{gao2020phygeonet} proposed a geometry-adaptive physics-informed CNN, PhyGeoNet, which embeds a pre-computed coordinate mapping into the classic CNN structure. Although the effectiveness of the PhyGeoNet has been demonstrated on simple irregular domains, it remains challenging for general complex geometries at large.

Motivated by existing challenges, we propose a novel discrete PINN framework to handle irregular domains with unstructured grids based on generalized convolution operations. Namely, the convolution operations are directly performed on unstructured mesh data, which can be seen as discrete non-euclidean manifolds, i.e., Graph. Moreover, the construction of PDE-informed graph convolutional network (GCN) structure is inspired by finite element (FE) method \cite{hughes2012finite, duarte1996h}, which is another classic numerical discretization technique that possess many advantages for physics-informed learning. First, thanks to a variational formulation (weak form) of PDE residuals, where Neumann boundary conditions can be naturally incorporated in the weak formulation of governing equations, the order of differential operators can be effectively reduced by integration by part and thus the learning complexity can be largely mitigated. Moreover, the massive amount of collocations points required by strong-form PINNs can be replaced by a relatively small amount of quadrature points, which could potentially reduce considerable training cost. The variational (weak) formulation has been recently developed for continuous PINNs and notable superiority has been shown over strong-form PINNs~\cite{weinan2018deep,samaniego2020energy,zang2020weak,kharazmi2019variational,kharazmi2021hp,khodayi2020varnet}. In these variational continuous PINNs, a point-wise fully-connected neural network is usually built as the trial basis, combined with polynomial test functions, to formulate the variational forms in Petrov-Galerkin fashion. Due to the black-box nature of the deep neural networks, accurate quadrature rules are difficult to construction, which leads to additional error associated with variational crimes. Moreover, the essential BCs cannot be imposed in a hard manner due to the point-wise formulation. The FEM-Net proposed by Yao et al.~\cite{yao2020fea} is a FE-based discrete PINN, where a FE-based convolution has been developed to build variational PDE residuals for CNNs. However, this method is under a linear assumption and still limited to rectangular domains due to the classic CNN backbone.

In this work, we proposed an innovative discrete PINN framework based on graph convolutional network and variational structure of PDE to solve forward and inverse PDEs in a unified manner. Specifically, the novel contributions are summarized as follows:
\begin{enumerate}[(a)]
 \item We introduce the graph convolution operation into physics-informed learning to fully leverage the power of FE-based discretization for irregular domains with unstructured meshes. Unlike the state-of-art discrete PINNs based on classic CNNs, the proposed approach does not need rasterization as it can directly handle unstructured mesh with simplex/quadrilateral elements as traditional FE solver does.
 \item A set of finite-dimensional polynomial basis functions are used to reconstruct the full-field predictions based on the output nodal solution graph in a Galerkin formulation, and thus, the search space can be significantly reduced to facilitate training. Moreover, since both test/trial function are based on standard polynomials, the variational integrals can be computed accurately using Gaussian quadrature.
 \item The proposed PINN is designed to exactly satisfy essential boundary conditions, avoiding penalty coefficient tuning in most PINNs with a soft BC enforcement.
 \item A new data assimilation scheme is proposed to strictly enforce observation data.
\end{enumerate}

\section{Methodology}
\label{sec:Methodology}
\subsection{Overview}
Consider a physical system in a bounded domain ($\Omega\subset\Rbb^d$) governed by a set of nonlinear, steady parameterized PDEs in the generic discretized form,
\begin{equation}
\Rbm(\Ubm(\mubold);\mubold)=0, 
\label{eqn:Residual}
\end{equation}
where $\mubold\in\Rbb^{N_{\mubold}}$ is the PDE parameter vector, $\Ubm:\Rbb^{N_{\mubold}}\rightarrow\Rbb^{N_{\Ubm}}$ is the discrete parameter-dependent state vector implicitly defined as the solution of (\ref{eqn:Residual}), and $\Rbm:\Rbb^{N_{\Ubm}}\times\Rbb^{N_{\mubold}}\rightarrow\Rbb^{N_{\Ubm}}$ represents the discretized PDE operator. The set of PDEs are subjected to  boundary conditions (BCs), which are defined on the boundary $\partial \Omega$ of the domain. In this work, we present an innovative physics-informed graph neural Galerkin network (PI-GGN) to establish a solution approach for such PDE-governed system in both forward and inverse settings. In the forward problem, we aim to obtain the solution $\Ubm$ given known BCs and parameters $\mubold$; as for the inverse setting, the system is solved when BCs and parameters $\mubold$ are partially known, whereas sparse observations of the state are available. In the proposed framework, a GCN is devised to learn nodal solutions of the state on a set of unstructured grids. The PDE residuals in the physics-informed loss function are reconstructed based on the continuous Galerkin method. Essential BCs of the system are imposed in a hard manner and additional data can be assimilated to solve the forward and inverse problems simultaneously. Each component of the proposed method will be detailed in the following subsections. 


\subsection{Graph convolutional neural network for unstructured data}
There has been growing interest in applying GCN for scientific machine learning problems because of its great flexibility in dealing with unstructured data. Excellent performance of the graph-based learning has been reported in modeling various computational mechanics problems through classic data-driven training~\cite{sanchez2018graph,battaglia2016interaction,maulik2019site,pfaff2020learning, sanchez2020learning,li2020multipole}. In general, by defining convolution operations for non-Euclidean space, GCNs generalize CNN-type constructions to graph data. The capability of modeling dependencies between nodes of a graph is the key that enables GCNs to handle unstructured mesh data with any arbitrary boundaries.
\begin{figure}[htp]
	\centering
	\includegraphics[width=0.7\textwidth, height=0.31111\textwidth]{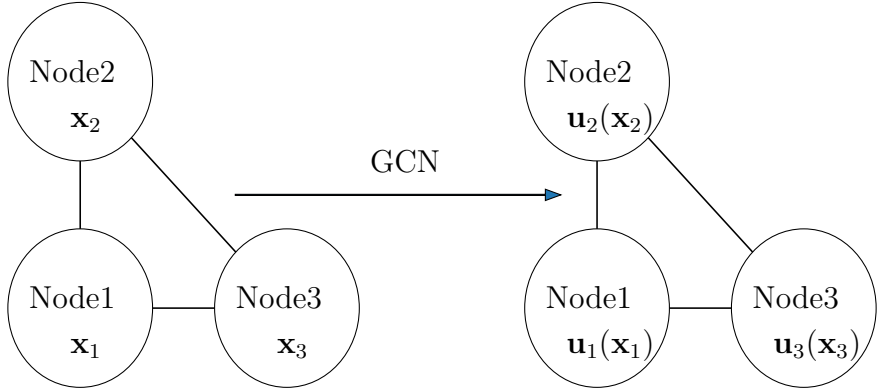}
	\caption{An example of a GCN, where the input/output graph has 3 nodes \& edges and the same adjacency matrix ($\Ncal(1)=\{2,3\}$, $\Ncal(2)=\{1,3\}$, $\Ncal(3)=\{1,2\}$). The input feature is the coordinate of each node ($\fbm_i^{(\mathrm{in})}=x_i$), while the output feature is the nodal solution vector ($\fbm_i^{(\mathrm{out})}=u_i(x_i)$).}
	\label{fig:graph_demo}
\end{figure}
As shown in Fig.~\ref{fig:graph_demo}, a graph consists of nodes and edges, where each node is defined by its feature vector $\fbm$ and the relation with other nodes are described by edges. The neighbor $\Ncal(\cdot)$ of a node refers to a set of adjacent nodes that are connected to that node via edges. Therefore, a mesh with unstructured grids and corresponding nodal PDE solutions can be naturally described as graphs. Similar to CNN-based discrete PINN~\cite{gao2020phygeonet}, a GCN is built to model the discretized solution fields $\Ubm(\bar\mubold) \approx \hat{\Ubm}(\Thetabold^*)$, where $\Thetabold^*$ are trained parameters of the GCN for graph convolutions for the parameter $\bar\mubold$. 
\begin{remark}
In general, the input feature vector of GCN can be any spatially varying field discretized by the mesh due to the universal approximation capacity of deep neural network. In this work, the GCN takes an input graph that each node is associated with its spatial coordinates of the mesh, and then outputs the discretized solutions fields as an out graph, where each node contains the corresponding nodal solution vector.
\end{remark}

Similar to CNNs, the output solution graph is obtained by applying multiple graph convolution operations on the input layer, sequentially updating nodal features via a message passing function, which can be written in a generic form,
\begin{equation}
	\fbm_i^{(l)}=\gamma^{(l)}(\fbm_i^{(l-1)},\square^{(l)}_{j\in\Ncal(i)}\Psi^{(l)}(\fbm_i^{(l-1)},\fbm_j^{(l-1)})),
\end{equation}
where $i$ denotes $i^{\mathrm{th}}$ node, $(l)$ denotes $l^{\mathrm{th}}$th layer, $\gamma$, $\Psi$ are differentiable non-linear functions, and $\square$ denotes a differentiable, permutation-invariant function (e.g., summation, mean, or maximum). The feature vectors are represented by $\fbm_i^{(l)}\in\Rbb^{N_{\fbm^{(l)}}}$ and $\fbm_i^{(l-1)}\in\Rbb^{N_{\fbm^{(l-1)}}}$, where $N_{\fbm^{(l-1)}}$ and $N_{\fbm^{(l)}}$ are feature dimensions in $(l-1)^{\mathrm{th}}$ and $l^{\mathrm{th}}$ layers, respectively. For implementation simplicity, all the nodal features are usually concatenated and flattened as a larger vector $\Xbm$. The information of edge connection is stored in a sparse matrix $\Abm$, known as the adjacency matrix. In this work, the GCN is constructed based on the Chebyshev spectral graph convolution operator~\cite{defferrard2016convolutional}, which is derived from the spectral convolution theorem~\cite{briandavies2001discrete}, where Chebyshev polynomials are introduced to avoid expensive eigen-decomposition. Specifically, the message passing function of Chebyshev graph convolution can be written as,
\begin{equation}
	\Xbm^l = \mathrm{ReLU}\left(\sum_{k=1}^{K}\Zbm^{(l-1,k)}\cdot\Thetabold^{(l-1,k)} + \bbm^{l-1}\right),
\end{equation}
where $\Thetabold^{(l-1,k)}$ are trainable parameters for the $k^\mathrm{th}$ basis in the $(l-1)^\mathrm{th}$ layer, $\bbm^{(l-1)}$ is an additive trainable bias vector, and the $k^\mathrm{th}$ basis $\Zbm^{(l-1,k)}$ is calculated recursively as follows,
\begin{equation}
\begin{split}
&\Zbm^{(l-1,1)} = \Xbm^{(l-1)},\\
&\Zbm^{(l-1,2)} = \hat{\Lbm}\cdot\Xbm^{(l-1)},\\
&\Zbm^{(l-1,k)} = 2\hat{\Lbm} \cdot \Zbm^{(l-1,k-1)} - \Zbm^{(l-1,k-2)},\\
\end{split}
\end{equation}
and
\begin{equation}
\begin{split}
&\hat{\Lbm}=\Lbm-\Ibm\\
&\Lbm =\Ibm-\Dbm^{-\frac{1}{2}}\Abm\Dbm^{-\frac{1}{2}}
\end{split}
\end{equation}
where $\Ibm$ is an identity matrix and $\Dbm$ represents the degree matrix of the graph. The Rectified Linear Unit (ReLU)~\cite{nair2010rectified} is chosen as the nonlinear activation function and polynomial order $K$ is set as $10$ in this work.


\subsection{Variational PDE-informed loss function}
The loss function is built based on the PDE residuals (Eq.~\ref{eqn:Residual}), such that the conservation laws are utilized to inform/drive the GCN training. The generic PDE for steady-state scenarios can be re-written as,
\begin{equation}
	\nabla\cdot F(u,\nabla u;\mubold) = S(u,\nabla u;\mubold)\quad\text{in }\Omega,
	\label{eqn:StrongFromVectorFrom}
\end{equation}
where $u:\Omega\rightarrow\Rbb^{N_c}$ is the solution variable, $F:\Rbb^{N_c}\rightarrow\Rbb^{N_c\times d}$ is the flux function, $S:\Rbb^{N_c}\rightarrow\Rbb^{N_c}$ is the source term, and $\nabla:=(\partial_{x_1},...,\partial_{x_d})$ denotes the gradient operator defined in the physical domain. Equation~\ref{eqn:StrongFromVectorFrom} can represent a wide range of static PDEs such as Poisson equation, linear elasticity equations, and Navier-Stokes equations. 

\subsubsection{Weak formulation of PDE residuals}
For continuous FC-PINNs, the derivative terms for constructing the PDE-informed loss function are obtained by AD in point-wise manner, and the FCNN as a continuous trial function searches an infinite-dimensional solution space. Therefore, the infinite search space over-complicates the non-convex optimization for the network training and a massive amount of collocation points are usually required. In this work, we use a piecewise polynomial basis to reduce the dimension of the search space and facilitate physics-informed training/convergence. Specifically, the conservation laws (Eq.~\ref{eqn:StrongFromVectorFrom}) are discretized based using a nodal continuous Galerkin method and the trial space $\Vcal_h^p$ is constructed by continuous piecewise polynomial basis functions
\begin{equation}
	\Vcal_h^p = \big\{ v\in[\Hcal^1(\Omega)]^{N_c}\;\big|\;v|_{K}\in[\Pcal_p(K)]^{N_c},\;\forall K\in\Ecal_h \big\},
\end{equation}
where $\Hcal^1(\Omega)$ represents Sobolev spaces where weak derivatives up to order one are square integrable, $\Pcal_p(K)$ is the space of polynomial functions of degree up to $p$ defined on the element $K$, and $\Ecal_h$ is the finite element mesh. The test space is set to be the same as the trial space $\Vcal_h^p$ and the solution $u_h\in\Vcal_h^p$ satisfies the weak formulation of the PDEs for any test function $\omega_h \in \Vcal_h^p$,
\begin{equation}
	\int_{\partial\Omega}\omega_h\cdot F(u_h,\nabla u_h;\mubold) n\,dS - \int_\Omega \nabla\omega_h : F(u_h,\nabla u_h;\mubold)\, dV = 
	\int_\Omega\omega_h\cdot S(u_h,\nabla u_h;\mubold)\,dV.
	\label{eqn:CGFormulation}
\end{equation}
We introduce a basis $\Phibold(x)\in\Rbb^{N_\Ubm\times N_c}$ for $\Vcal_h^p$ to express the test variables as
$\omega_h(x) = \Phibold(x)^T \tilde\Wbm$, where $\tilde\Wbm\in\Rbb^{N_\Ubm}$ are the coefficients of the test
variable in the basis, which leads to an equivalent version of the Galerkin form
\begin{equation}
\int_{\partial\Omega}\Phibold \cdot F\Big(u_h,\nabla u_h;\mubold\Big) n\,dS
-\int_\Omega \nabla\Phibold : F\Big(u_h,\nabla u_h;\mubold\Big) \,dV -
\int_\Omega\Phibold\cdot S\Big(u_h,\nabla u_h;\mubold\Big)\,dV=0.
\label{eqn:weakFormContinuous}
\end{equation}
using arbitrariness of the test function coefficients. We convert this to residual form by introducing $\{(\beta_i^v,\tilde{x}_i^v)\}^{N_{qv}}_{i=1}\}$ and $\{(\beta_i^s,\tilde{x}_i^s)\}^{N_{qs}}_{i=1}\}$
as the quadrature weights and points for integrals over $\Omega$ and $\partial\Omega$, respectively, to define the residual as
\begin{equation}
\begin{split}
\Rbm(\tilde\Ubm;\mubold)=&\sum_{i=1}^{N_{qs}} \beta^s_i \Phibold(\tilde{x}^s_i) \cdot F\Big(\tilde{u}_h(\tilde{x}_i^s;\tilde\Ubm), \nabla \tilde{u}_h(\tilde{x}_i^s;\tilde\Ubm);\mubold\Big) n - \\
&\sum_{i=1}^{N_{qv}} \beta^v_i \nabla\Phibold(\tilde{x}^v_i) : F\Big(\tilde{u}_h(\tilde{x}_i^v;\tilde\Ubm), \nabla \tilde{u}_h(\tilde{x}_i^v;\tilde\Ubm);\mubold\Big)  -\\
&\sum_{i=1}^{N_{qv}} \beta^v_i \Phibold(\tilde{x}^v_i) \cdot S\Big(\tilde{u}_h(\tilde{x}_i^v;\tilde\Ubm), \nabla \tilde{u}_h(\tilde{x}_i^v;\tilde\Ubm);\mubold\Big),
\end{split}
\label{eqn:residualFull}
\end{equation}
where $\tilde{u}_h : \Omega \times \Rbb^{N_\Ubm} \rightarrow \Rbb^{N_c}$ is the continuous representation in $\Vcal_h^p$ of the discrete state vector, i.e.,
\begin{equation}
	\tilde{u}_h(x;\tilde\Ubm) = \Phibold(x)^T\tilde\Ubm.
	\label{eqn:GalerkinExp}
\end{equation}

The surface and volume quadrature coefficients ($\beta^s$ and $\beta^v$) are stored as constant tensors and remain unchanged during the network training. The matrix of basis function $\Phibold$ are obtained on the limited amount of quadrature points and can be pre-computed as constant tensors ($\Phibold(\tilde{x}^v), \Phibold(\tilde{x}^s),\nabla\Phibold(\tilde{x}^v),\nabla\Phibold(\tilde{x}^s)$). The variational formulation of the PDE residual (eq.~\ref{eqn:residualFull}) will be used to define the physics-informed loss function for the GCN. Namely, the nodal solution vector $\tilde{\Ubm}$ will be learned by GCN as the output graph $\hat{\Ubm}(\Thetabold)$, which takes the coordinates ($\chibold$) as the input graph. When the PDE parameters $\mubold$ are unknown, they can be treated as trainable parameters, being updated along with network parameters $\Thetabold$. Both the flux and source functions ($F,S$) are differentiable functions, where gradient information can be propagated from the outputs to their inputs. Table~\ref{tab:treatment} summarizes these notations.

\begin{table}[htp]
	\small
	\begin{center}
		\begin{tabular}{ |c|c|c|c|  } 
			\hline
			Notations& Description & \multicolumn{2}{c|}{Treatment in PI-GGN}   \\
			\hline
			$\mubold$& PDE parameter &Constant (if known)& Trainable (if unknown)\\
			\hline
			$\beta_i^v$, $\beta_i^s$ & Quadrature weights  &\multicolumn{2}{c|}{Constant tensors}\\
			\hline
			$\Phibold(\cdot)$& Basis function& \multicolumn{2}{c|}{Constant tensors}   \\
			\hline
			$F$, $S$ &Flux and source functions &\multicolumn{2}{c|}{Differentiable functions} \\
			\hline
			$\hat{\Ubm}$ &Nodal solution & \multicolumn{2}{c|}{Output graph of the GCN}\\
			\hline
			$\chibold$ &Nodal coordinates & \multicolumn{2}{c|}{Input graph of the GCN}\\
			\hline
		\end{tabular}
		\caption{Summary of notation}
		\label{tab:treatment}
	\end{center}
\end{table}

\subsubsection{Essential boundary conditions enforcement} 
We apply static condensation to (\ref{eqn:residualFull}) by restricting to the unconstrained degrees of freedom, e.g., degrees of freedom away from essential BCs, to yield
\begin{equation}
 \Rbm_u(\Ubm_u(\mubold), \Ubm_e; \mubold) = 0,
\end{equation}
where $\Ubm_e$ are the known value of the essential boundary conditions and $\Ubm_u(\mubold)$ are the indices of $\Ubm(\mubold)$ corresponding to the
unconstrained degrees of freedom. In the neural network setting, we enforce the essential boundary conditions strongly by partitioning the degrees of freedom into
unconstrained (unknown) and constrained (known) degrees of freedom as $\hat\Ubm(\Thetabold) = (\hat\Ubm_u(\Thetabold)^T,\hat\Ubm_c^T)^T$ and defining the constrained
degrees of freedom using the known value of the essential BCs, i.e., $\hat\Ubm_c = \Ubm_e$, and the unconstrained degrees of freedom by minimizing the physics-informed
loss function
\begin{equation}
\Lcal_{\mathrm{f}}(\Thetabold;\mubold)=\left\|\Rbm_u\left(\hat{\Ubm}_u(\Thetabold),\Ubm_e;\mubold\right)\right\|_2.
\label{eqn:forwardResidiual}
\end{equation}
In this formulation, the essential boundary condition will be satisfied automatically by construction, which is in contrast to continuous FC-PINN that defines the FCNN as a point-wise solution function, posing challenges in hard boundary enforcement.

\subsection{Unifying forward and inverse solutions}
The GCN can be trained based on the physics-informed loss function defined in Eq.~\ref{eqn:forwardResidiual} by solving the following optimization problem without labels, 
\begin{equation}
\label{eqn:pdetraining}
\Thetabold^* = \underset{\Thetabold}{\arg\min}~\Lcal_{\mathrm{f}}(\Thetabold;\bar\mubold)
\end{equation}
where $\Thetabold^*$ denotes optimal network parameters and $\bar\mubold$ are the known PDE parameters; the GCN is then used to solve a forward PDE (\emph{forward solution}).
However, in many cases, some physical parameters such as material properties, inlet velocity, and Reynolds number, are not available, while sparse observation data (labels) $\Ubm_o$ can be obtained, which can be assimilated to infer the unknown parameters (\emph{inverse solution}). In previous PINN approaches, the inverse problem can be solved by assimilating data  $\Ubm_o$ in a soft manner, where the physics-informed loss is augmented by a data loss component. Namely, the following optimization is formulated,
\begin{equation}
\label{eqn:softInverse}
(\Thetabold^*, \mubold^*) = \underset{\Thetabold, \mubold}{\arg\min}~\Lcal_{\mathrm{f}}(\Thetabold;\mubold) + \lambda\underbrace{ \left\| \mathcal{\Fbm}^{s2o}\left(\hat{\Ubm}(\Thetabold)\right) -\Ubm_o \right\|_2}_{\text{data loss:}\ \mathcal{L}^d},
\end{equation}
where $\mathcal{\Fbm}^{s2o}$ represents the state-to-observable map and $\lambda$ is the penalty parameter. Properly tuning the penalty weight $\lambda$ is critical to the convergence, which is, however, challenging and often conducted empirically~\cite{wang2020understanding}. Here we introduce a novel approach to assimilate observation data and infer unknown parameters without the need of hyperparameter tuning. Specifically, the observation data are strictly imposed by constructing the GCN output as
 \begin{equation}
 \label{eqn:hard}
 \mathcal{\Fbm}^{s2o}\left(\hat{\Ubm}(\Thetabold)\right) =  \Ubm_o
 \end{equation}
Therefore, unknown parameters $\mubold$ and boundary conditions $\hat{\Ubm}_u$ can be obtained along with the PDE solutions $\hat{\Ubm}_u$ simultaneously by solving the following constrained optimization problem,
\begin{equation}
\label{eqn:hardtrain}
(\Thetabold^*, \mubold^*) = \underset{\Thetabold, \mubold}{\arg\min} ~\Lcal_{\mathrm{f}}(\Thetabold;\mubold), \quad \text{subject to:} \quad
 \mathcal{\Fbm}^{s2o}\left(\hat{\Ubm}(\Thetabold)\right)  = \Ubm_o.
\end{equation}

\section{Numerical experiments}
\label{sec:NumericalExperiments}
We demonstrate the proposed physics-informed graph Galerkin neural network (PI-GGN) on a variety of computational mechanics problems in both forward and inverse settings. Specifically, Poisson equations, linear elasticity equations, and Navier-Stokes equations with known or unknown BCs/parameters are investigated here to demonstrate the effectiveness of the proposed method. Moreover, we also compare two different ways of assimilating sparse observation data and show the advantage of strictly enforcing data for the parameter/field inversion. For all cases, the GCN architecture remains the same, where the dimensions of node vector in hidden graph lays are fixed as $[32, 64, 128, 256, 128, 64, 32]$. The relative error metric $e$ is defined as,
\begin{equation}
e = \frac{||\hat{\Ubm}(\Thetabold^*)-\Ubm(\bar\mubold)||_2}{||\Ubm(\bar\mubold)||_2},
\end{equation}
where $\Thetabold^*$ is the optimal training parameters computed for the parameter configuration $\bar\mubold$.
\subsection{Poisson equation}
We start from a 2-D homogeneous Poisson equation,
\begin{equation}
\begin{split}
	&f+\Delta u=0\quad\text{in }\Omega,\quad \\
	&u=0\quad\text{on }\partial\Omega,
\end{split}
\label{eqn:poi}
\end{equation}
where $u$ is the primary variable, $f$ is the source term, and $\Delta$ denotes the Laplacian operator.

\subsubsection{Forward solution of diffusion field}
\begin{figure}[htp]
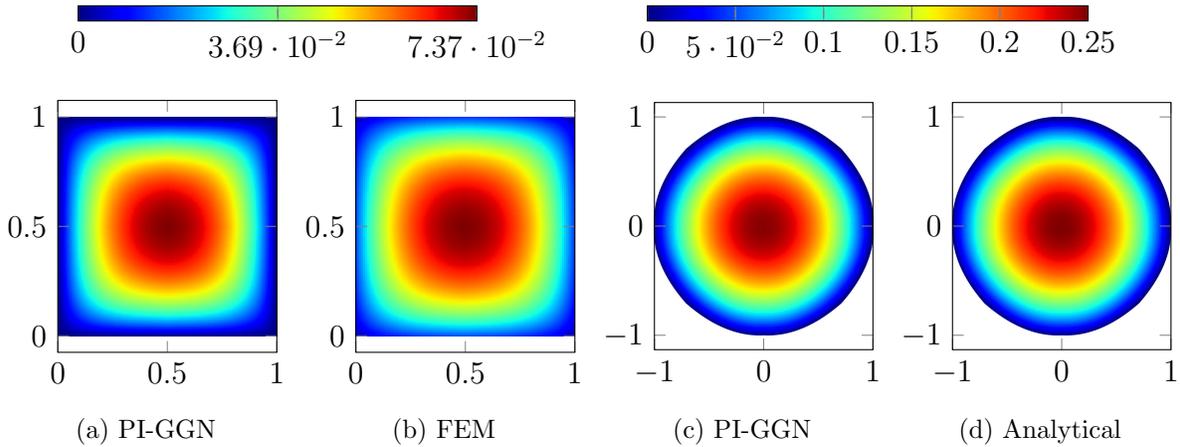

	\centering
	\includegraphics[width=0.4\textwidth,height=0.05\textwidth]{colorbar_possion_square.tikz}\hspace{2em}
	\includegraphics[width=0.4\textwidth,height=0.05\textwidth]{colorbar_possion_circle.tikz}
	\vfill
	\centering
	\subfloat[PI-GGN]
	{\includegraphics[width=0.24\textwidth,height=0.24\textwidth]{gcnn_possion_square_nocolorbar.tikz}}
	\subfloat[FEM]
	{\includegraphics[width=0.24\textwidth,height=0.24\textwidth]{FEM_possion_square_nocolorbar.tikz}}
	\subfloat[PI-GGN]
	{\includegraphics[width=0.24\textwidth,height=0.24\textwidth]{gcnn_possion_circle_nocolorbar.tikz}}
	\subfloat[Analytical]
	{\includegraphics[width=0.24\textwidth,height=0.24\textwidth]{exact_possion_circle_nocolorbar.tikz}}
	\caption{PI-GGN forward solutions of the diffusion field $u$ on the (a) square and (c) circular disks, compared against corresponding FEM or analytical solutions, where the relative prediction error of the PI-GGN is $e = 5\times10^{-3}$ on the square domain and $e = 5\times10^{-4}$ on the circular disk, respectively.}
	\label{fig:PossionSquareAndDisk}
\end{figure}

We first consider the forward problem, where the source term $f$ is given ($f = 1$) over a unit square domain (Figs.~\ref{fig:PossionSquareAndDisk}a and~\ref{fig:PossionSquareAndDisk}b). Four quadrilateral elements are used to discretize the domain with $3^{\mathrm{rd}}$order of polynomial basis for solution and domain transformation. 
As a result, the total nodal points of the graph is 49, which is much lower than the total number of collocation points for a typical point-wise FC-PINN. The contour of the PI-GGN prediction is in a good agreement with the FEM reference, and the relative error is $e = 0.5\%$, though slight under-estimation near the boundary is observed. In Fig.~\ref{fig:PossionSquareAndDisk}c, the same PDE is solved on a unit circular domain, where the analytical solution exists (Fig.~\ref{fig:PossionSquareAndDisk}d),
\begin{equation}
\pvar(x,y)=\frac{1-x^2-y^2}{4}.
\end{equation}
In PI-GGN, the number of elements remains the same, while the order of the polynomial basis is set as two and thus 25 nodal points are used to construct the graph. We can see the PI-GGN forward solution is almost identical to the analytical reference and the relative prediction error $e$ is only $0.05\%$. This simple test case demonstrates that the graph-based discrete PINN can easily handle non-rectangular domains with unstructured meshes, which have posed challenges for standard FD-based CNN architectures, where special treatment such as rasterization or coordinate transformation is required~\cite{gao2020phygeonet}, complicating the implementation and convergence. 



\subsubsection{Inverse solution of unknown source term}
\begin{figure}[htp]
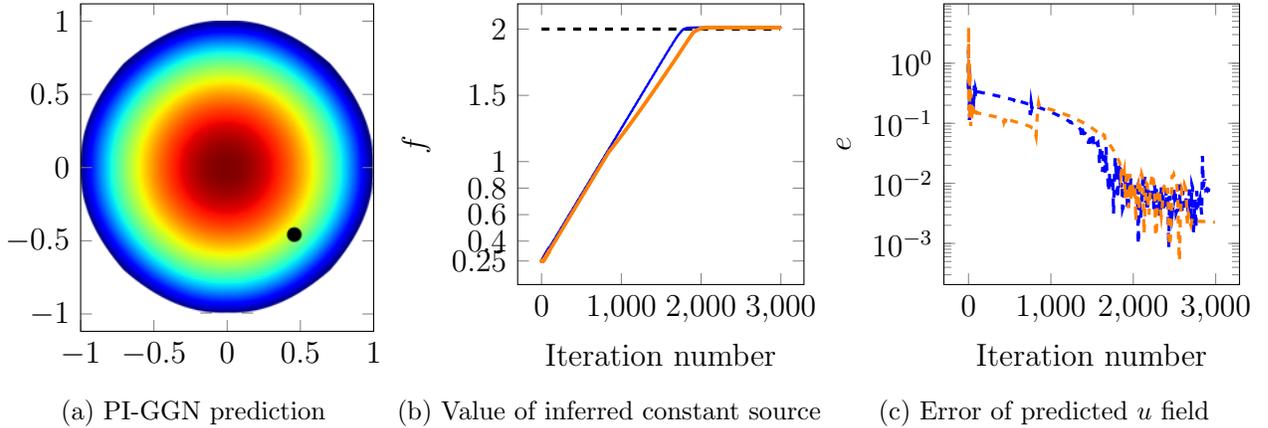

	\centering
	\subfloat[PI-GGN prediction]
	{\includegraphics[width=0.3\textwidth,height=0.3\textwidth]{posobslocation.tikz}}
	\subfloat[Value of inferred constant source]
	{\includegraphics[width=0.35\textwidth,height=0.3\textwidth]{PossionInfer.tikz}}
	\subfloat[Error of predicted $u$ field]
	{\includegraphics[width=0.35\textwidth,height=0.3\textwidth]{PossionInferSolutionError.tikz}}
	\caption{PI-GGN inverse solutions of the source term $f$ by assimilating observed diffusion data (black dots) using (1) penalty method (\ref{fig:SoftPosInvVal}) and (2)  hard enforcement (\ref{fig:HardPosInvVal}), compared against the ground truth (\ref{fig:TruePosInvVal}), where the error of field prediction by soft (\ref{fig:SoftPosInvEr}) and hard  (\ref{fig:HardPosInvEr}) data assimilation are presented.}
	\label{fig:posinvresult}
\end{figure}

The real power of the PI-GGN is to solve the forward and inverse problems simultaneously by assimilating additional state observations. For example, when the source term is not given, the PI-GGN is able to assimilate sparse data to solve the diffusion field and meanwhile infer the unknown source term in a unified manner. Here we assume the constant source term $f = 2$ is unknown and observation of $u$ is available only at one point as shown in Fig.~\ref{fig:posinvresult}a. We use two ways to assimilate the data and solve the inverse problem: one is to assimilate data by adding a data loss as a penalty term with (Eq.~\ref{eqn:softInverse}) with the hyper-parameter chosen as $\lambda = 1000$, and the other is to assimilate data strictly based on Eq.~\ref{eqn:hardtrain}. As shown in Fig.~\ref{fig:posinvresult}b, the inferred source terms from both approaches converge to the ground truth and the forward solution of the $u$ field is also obtained \emph{simultaneously}. Overall, the prediction errors of the unknown source term and diffusion field are less than $1\%$.

\subsection{Linear elasticity equations}
Next, we consider problems governed by linear elasticity equations,
\begin{equation}
\begin{split}
\nabla \cdot \sigma = 0& \quad\text{in }\Omega, \\
\sigma \cdot n = t& \quad\text{on } \partial\Omega^N, \\
u = u^D&\quad\text{on }\partial\Omega^D,
\end{split}
\label{eqn:linearelasticity}
\end{equation}
where $u:\Omega\rightarrow\Rbb^d$ is the displacement vector, $\sigma: \Omega\rightarrow\Rbb^{d\times d}$ is the stress tensor defined as $\sigma_{ij} = \lambda u_{kk} \delta_{ij} + \mu (u_{i,j}+u_{j,i})$, $n:\partial\Omega\rightarrow\Rbb^d$ is the unit normal vector on the boundary, $t:\partial\Omega^N\rightarrow\Rbb^d$ is the applied traction force, $u^D:\partial\Omega^D \rightarrow \Rbb^d$ is the essential boundary condition, and $\lambda$ and $\mu$ are the constant Lam\'e parameters. For each variable component $\pvar_i$, a sub-GCN is constructed for the prediction.


\subsubsection{Forward solution of displacement field}
\begin{figure}[htp]
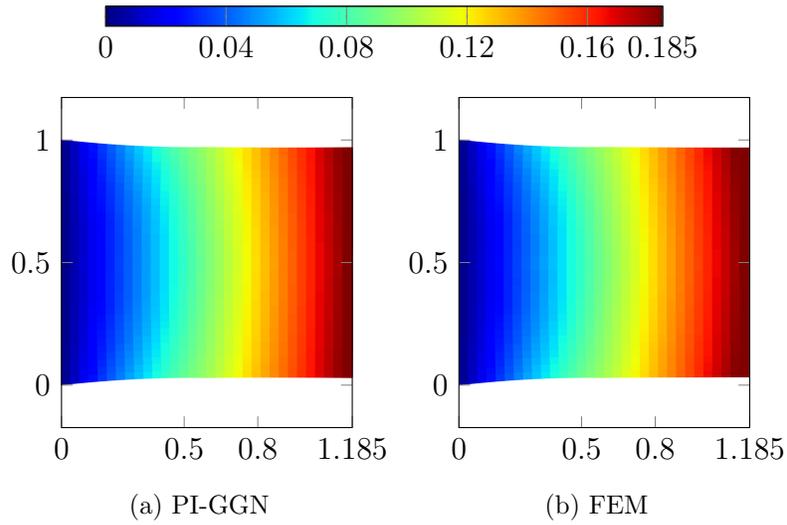

	\centering
	\includegraphics[width=0.5\textwidth,height=0.05\textwidth]{2Dlinear_elasticity_square_colorbar.tikz}
	\centering
	\subfloat[PI-GGN]
	{\includegraphics[width=0.32\textwidth,height=0.3\textwidth]{gcnn_2dlinearelasticity_square.tikz}}
	\subfloat[FEM]
	{\includegraphics[width=0.32\textwidth,height=0.3\textwidth]{fem_2dlinearelasticity_square.tikz}}
	\caption{PI-GGN forward solutions of the displacement field $u$, compared against corresponding FEM reference, where the relative prediction error of the PI-GGN is $e = 1\times10^{-2}$.}
	\label{fig:linearelasticitySquare}
\end{figure}
First, we solve the forward problem in a unit square domain. To discretize this domain, four quadrilateral elements are used, and the order of polynomial basis for solution and domain transformation is set as two, resulting in a $25$-nodal graph. The Lamé parameters are set as $\lambda=1$ and $\mu=1$. The essential boundary condition $u = [0, 0]$ is prescribed on the left side ($x=0$) and the natural boundary condition $t=[0.5, 0]$ is imposed on the right side. Fig.~\ref{fig:linearelasticitySquare} shows that the PI-GGN forward solution of the displacement field agrees with the FEM reference very well.

\begin{figure}[htp]
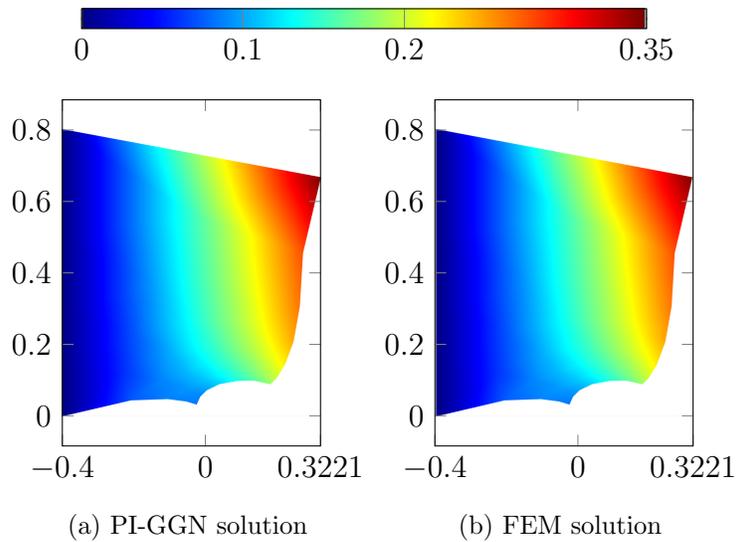

	\centering
	\includegraphics[width=0.5\textwidth,height=0.05\textwidth]{2d_colorbar_simplex.tikz}
	\centering
	\subfloat[PI-GGN solution]
	{\includegraphics[width=0.3\textwidth,height=0.315\textwidth]{gcnn_2d_simplex.tikz}}
	\subfloat[FEM solution]
	{\includegraphics[width=0.3\textwidth,height=0.315\textwidth]{fem_2d_simplex.tikz}}
	\caption{PI-GGN forward solutions of the displacement field $u$, compared against corresponding FEM reference, where the relative prediction error of the PI-GGN is $e = 5\times10^{-3}$.}
	\label{fig:linearelasticitySimplex}
\end{figure}
Then we investigate a irregular domain, a rectangular with a notch, where same Lamé parameters are specified. The domain is discretized by $55$ simplex elements with 1st order polynomial basis for the solution and domain transformation. The essential boundary conditions $u^D=[0,0]$ is imposed at the left boundary side $x=-0.4$ and the natural boundary condition $t_1=0.5$ is prescribed at the right boundary side. As mentioned above, no special treatment is needed for PI-GGN to handle irregular geometry with simplex mesh. Fig.~\ref{fig:linearelasticitySimplex} shows that the forward solution by PI-GGN are very accurate compared to the FEM reference. 

\begin{figure}[htp]
	\centering
	\includegraphics[width=0.5\textwidth,height=0.05\textwidth]{3d_colorBarModify.tikz}
	\vfill
	\centering
	\subfloat[PI-GNN]
	{\includegraphics[width=0.4\textwidth]{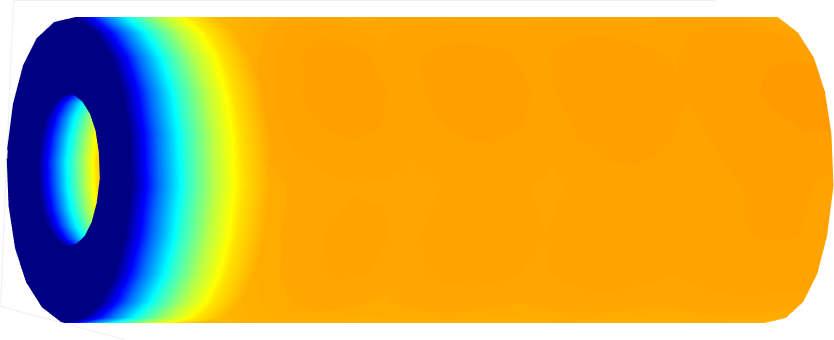}}
	\subfloat[FEM]
	{\includegraphics[width=0.4\textwidth]{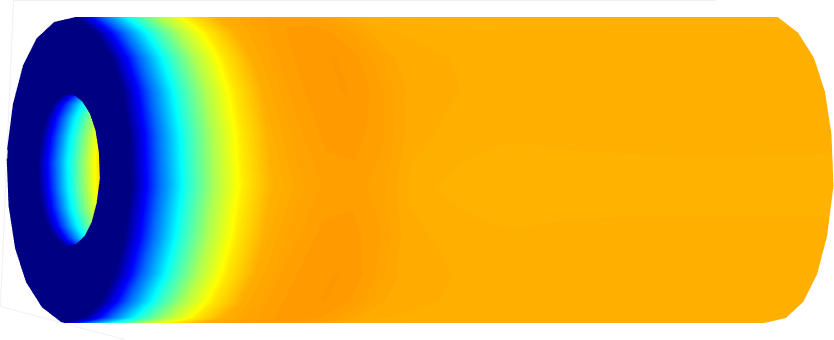}}
	\caption{PI-GGN forward solutions of the displacement field $u$, compared against corresponding FEM reference, where the relative prediction error of the PI-GGN is $e = 5\times10^{-2}$.}
	\label{fig:3dContour}
\end{figure}

Lastly, we consider a 3-D domain. Specifically, the deformation of a 3-D hollow cylinder is solved by the PI-GGN. The essential boundary conditions $u^D=[0, 0, 0]$ are imposed at the left surface, the Neumann boundary conditions $t=-n$ are prescribed at the inner surface of the cylinder ($x^2+y^2=1$), and $t=[0, 0, -0.25]$ are imposed at the right surface. The second order polynomial basis is used and the number of hexahedral element is $40$ with $440$ nodal points. The Lam\'e parameters are set as $\lambda=0.73$ and $\mu=0.376$. The forward solution of the displacement by PI-GGN agree with the FEM reference reasonably well, though PI-GGN slightly over-predicts the displacement of the right end of the cylinder (Fig.~\ref{fig:3dContour}). 
 
\subsubsection{Inverse solution of unknown material properties}
Next, we solve an inverse problem governed by the linear elasticity equations (Eq.~\ref{eqn:linearelasticity}). The Lam\'e parameters ($\lambda$ and $\mu$) are assumed to be unknown, whose true values are set as  $\lambda=\mu=1$.
\begin{figure}[htp]
	\centering
	\subfloat[PI-GGN prediction]
	{\includegraphics[width=0.4\textwidth,height=0.3\textwidth]{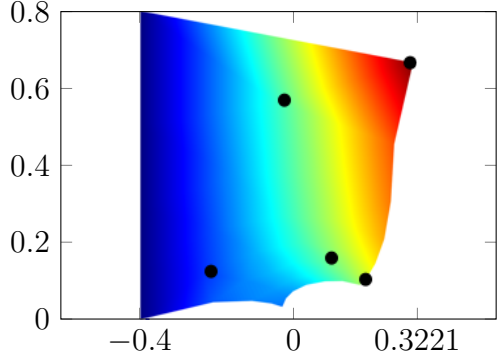}}
	\hspace{0.1\textwidth}
	\subfloat[Errors of predicted $u$ field]
	{\includegraphics[width=0.4\textwidth,height=0.285\textwidth]{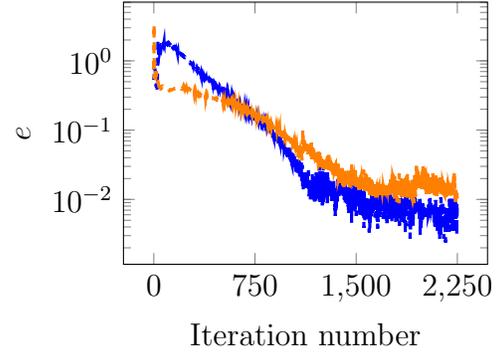}}
	\vfill
	\subfloat[$\lambda$]
	{\includegraphics[width=0.4\textwidth,height=0.3\textwidth]{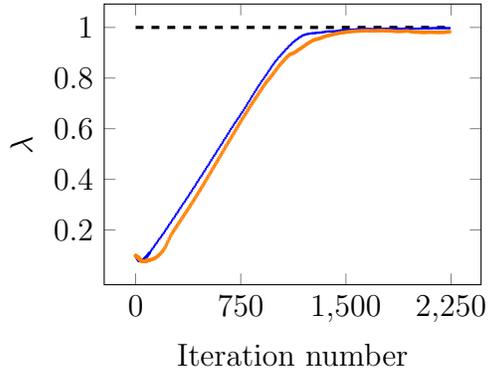}}
	\hspace{0.1\textwidth}
	\subfloat[$\mu$]
	{\includegraphics[width=0.4\textwidth,height=0.3\textwidth]{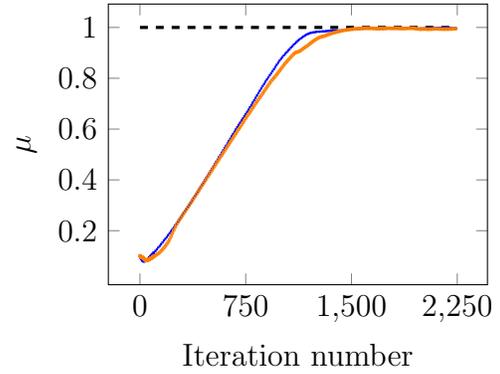}}
	
	\caption{PI-GGN inverse solutions of the Lam\'e parameters by assimilating observed displacement data (black dots) using (1) penalty method (\ref{fig:SoftLinelaInvVal}) and (2)  hard enforcement approach (\ref{fig:HardLinelaInvVal}), compared against the ground truth (\ref{fig:TrueLinelaInvVal}), where the error of field prediction by soft (\ref{fig:SoftLinelaInvEr}) and hard (\ref{fig:HardLinelaInvEr}) data assimilation are presented.}
	\label{fig:linelainv}
\end{figure}
The displacement field is observed at five randomly selected points shown in Fig.~\ref{fig:linelainv}(a). The entire field of the displacement is obtained via PI-GGN and the Lam\'e parameters can be inferred accurately as well (Figs.~\ref{fig:linelainv}c and ~\ref{fig:linelainv}d). The relative error of the PI-GGN predicted displacement field is $0.005$ by assimilating data in a hard manner, which is slightly lower than that of using penalty method ($e = 0.01$), shown in Fig~\ref{fig:linelainv}b.  

\subsection{Naiver-Stokes equations}
In the last test case, we study forward and inverse problems governed by the static incompressible Navier-Stokes (NS) equations, which is more challenging due to its strong nonlinearity. The steady NS equations model the viscous fluid flow with a constant density, which can be expressed as,
\begin{equation}
	\begin{split}
	(v \cdot\nabla) v-\nu\Delta v+\nabla p =0,\quad\nabla \cdot v=0\quad\text{in }\Omega,\\
	v = v^D\quad\text{on }\partial\Omega^D,\\
	\nu(n \cdot\nabla)v-pn=0\quad\text{on }\partial\Omega^N,
	\end{split}
\end{equation}
where $v:\Omega\rightarrow\Rbb^d$ is the velocity vector, $p:\Omega\rightarrow\Rbb$ is the pressure, $\nu$ is the the viscosity of the fluid, and $n:\partial\Omega\rightarrow\Rbb^d$ is the unit outward normal vector to the boundary. The solution variable vector is denoted by $u=[v_1,v_2,p]$. The viscosity is set as $\nu=0.01$. For stability reasons, a mixed element approximation is adopted~\cite{letallec1980mixed}. A separate sub-net is constructed for prediction of each of the solution variables $v_1$, $v_2$ and $p$.

\subsubsection{Forward solution of velocity and pressure fields}
\begin{figure}[htp]
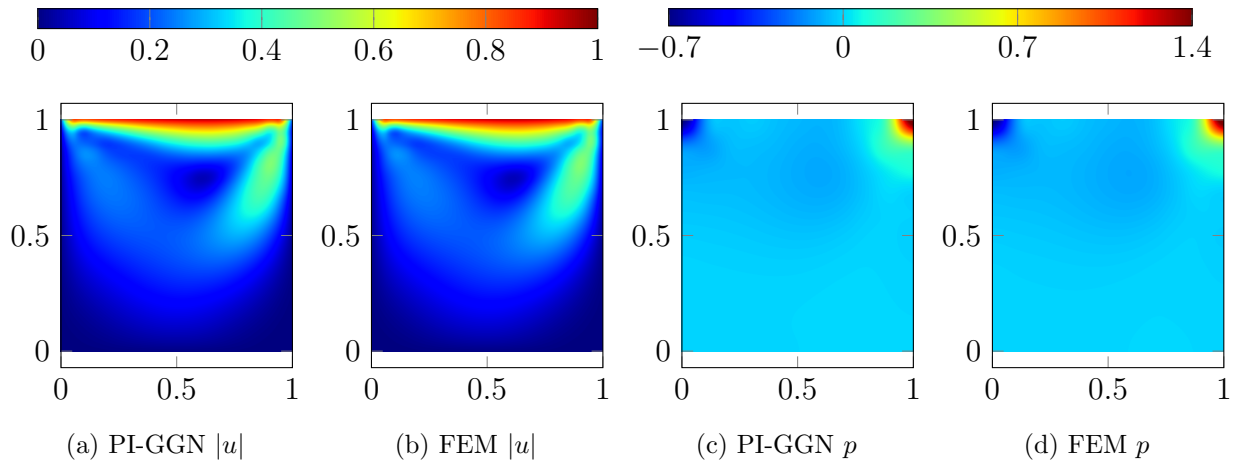

	\centering
	\includegraphics[width=0.48\textwidth,height=0.05\textwidth]{Umag_Colorbar_NS_LID.tikz}
	\includegraphics[width=0.48\textwidth,height=0.05\textwidth]{P_Colrobar_NS_Lid.tikz}
	\subfloat[PI-GGN $|u|$]
	{\includegraphics[width=0.25\textwidth,height=0.25\textwidth]{GCNN_U_LID.tikz}}
	\subfloat[FEM $|u|$]
	{\includegraphics[width=0.25\textwidth,height=0.25\textwidth]{FEM_U_LID.tikz}}	
	\centering
	\subfloat[PI-GGN $p$]
	{\includegraphics[width=0.25\textwidth,height=0.25\textwidth]{GCNN_P_LID.tikz}}
	\subfloat[FEM $p$]
	{\includegraphics[width=0.25\textwidth,height=0.25\textwidth]{FEM_P_LID.tikz}}

	\caption{PI-GGN forward solutions of the velocity magnitude and pressure fields, compared against corresponding FEM reference, where the relative errors is $8.7\times 10^{-3}$ for the velocity prediction and $1.95\times 10^{-2}$ for the pressure prediction.}
	\label{fig:NSforwardLidContour}
\end{figure}
First, we test the proposed approach on a classic flow problem, lid-driven cavity flow, defined on a square domain. The lid is placed on the top edge and moves rightward ($v_1 = 1, v_2 = 0$). The remaining three edges are set as no-slip walls  ($v_1=v_2=0$). The domain is discretized by $100$ quadrilateral elements. The numbers of collocation points for velocity and pressure fields are $441$ and $121$, respectively. The contours of forward solutions of velocity and pressure by PI-GGN is in a good agreement with the corresponding FEM reference, as shown in Fig.~\ref{fig:NSforwardLidContour}. The relative prediction errors are less than $1\%$. It is worth noting that over $10000$ collocation points were used to achieve same level of accuracy for AD-based FC-PINN~\cite{wang2020understanding,jagtap2020conservative}. 

\begin{figure}[htp]
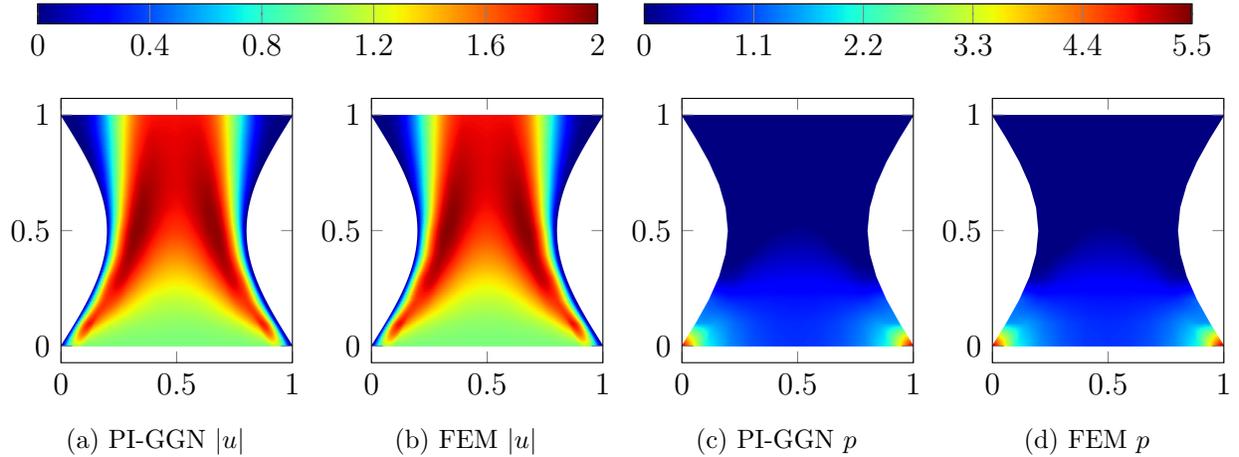

	\centering
	\includegraphics[width=0.48\textwidth,height=0.05\textwidth]{STE_v_colorbar.tikz}
	\includegraphics[width=0.48\textwidth,height=0.05\textwidth]{STE_p_colorbar.tikz}
	\subfloat[PI-GGN $|u|$]
	{\includegraphics[width=0.25\textwidth,height=0.25\textwidth]{GCNN_U_STE.tikz}}
	\subfloat[FEM $|u|$]
	{\includegraphics[width=0.25\textwidth,height=0.25\textwidth]{FEM_U_STE.tikz}}
	\centering
	\subfloat[PI-GGN $p$]
	{\includegraphics[width=0.25\textwidth,height=0.25\textwidth]{GCNN_P_STE.tikz}}
	\subfloat[FEM $p$]
	{\includegraphics[width=0.25\textwidth,height=0.25\textwidth]{FEM_P_STE.tikz}}
	\caption{PI-GGN forward solutions of the velocity magnitude and pressure fields, compared against corresponding FEM reference, where the relative errors is  $4.4\times 10^{-3}$ for the velocity prediction and $1.8\times 10^{-2}$ for the pressure prediction.}
	\label{fig:NSforwardStenosisContour}
\end{figure}
We also test the PI-GGN on solving the fluid flow in an idealized stenosis, where the inlet velocity is set as $v^D=[0,1]$ at the bottom ($y=0$) and no-traction boundary condition is prescribed at the outlet on the top ($y=0$). The same finite element setting is used as the lid-driven cavity problem. Similarly, both the velocity and pressure fields can be accurately solved and the PI-GGN predictions agree with the FEM reference well.

\subsubsection{Inverse solution of unknown inlet velocity field and unobserved pressure field}
Lastly, we consider an inverse problem governed by the NS equations.
\begin{figure}[htp]
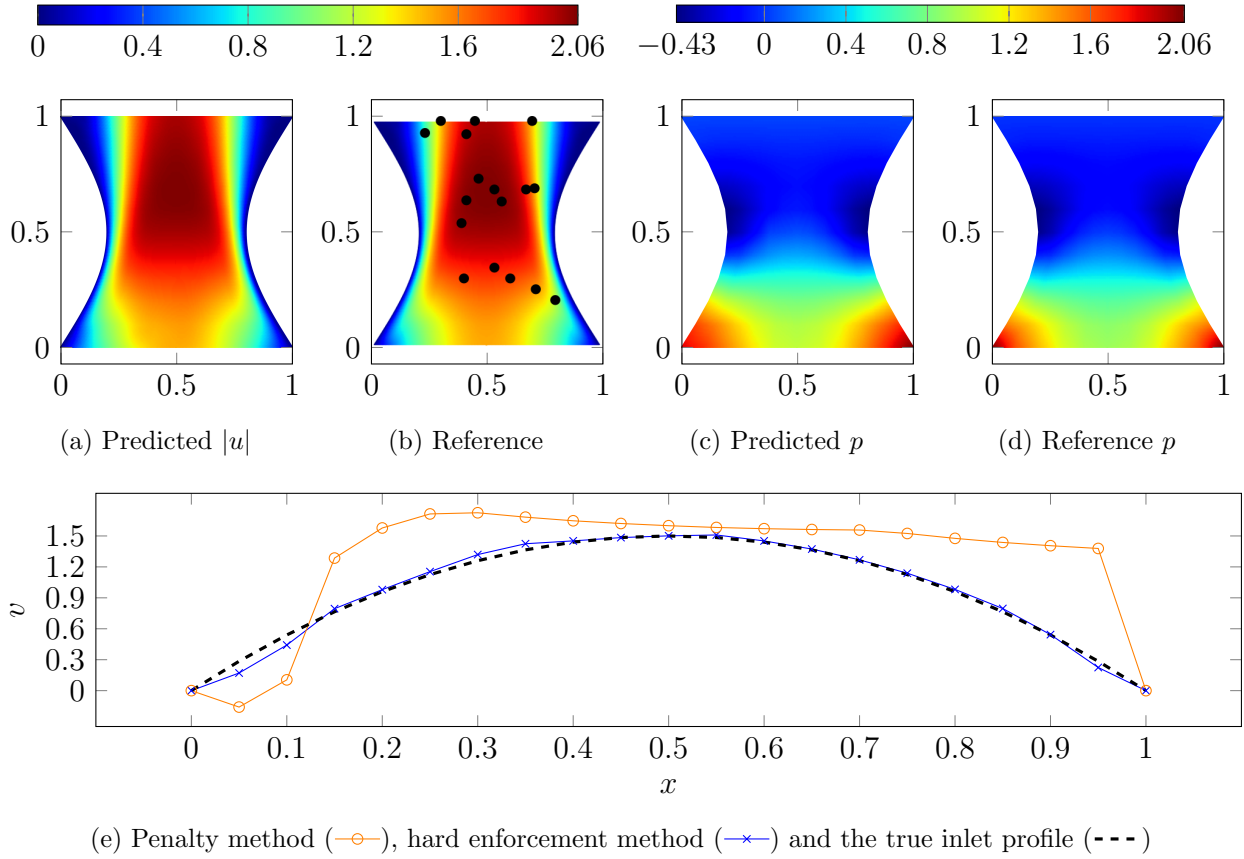

	\centering
	\includegraphics[width=0.48\textwidth,height=0.05\textwidth]{InvNSUbar.tikz}
	\includegraphics[width=0.48\textwidth,height=0.05\textwidth]{InvNSPbar.tikz}
	\subfloat[Predicted $|u|$]
	{\includegraphics[width=0.25\textwidth,height=0.25\textwidth]{InvNSGCNNU.tikz}}
	\subfloat[Reference]
	{\includegraphics[width=0.25\textwidth,height=0.25\textwidth]{NSOBS1.tikz}}
	\subfloat[Predicted $p$]
	{\includegraphics[width=0.25\textwidth,height=0.25\textwidth]{InvNSGCNNP.tikz}}	
	\subfloat[Reference $p$]
	{\includegraphics[width=0.25\textwidth,height=0.25\textwidth]{NSInvPressure.tikz}}
	\centering
	\vfill
	\subfloat[Penalty method (\ref{fig:SoftInvVal}), hard enforcement method (\ref{fig:HardInvVal}) and the true inlet profile (\ref{fig:TrueInvVal})]
	{\includegraphics[width=1.0\textwidth,height=0.25\textwidth]{NSInferInletProfile1.tikz}}
	\caption{PI-GGN inverse solutions of the inlet velocity field by assimilating observed velocity data at 19 randomly selected points. The relative error of the Inferred inlet profile is $e = 0.4$ by the soft penalty method while $e = 0.04$ by hard enforcement approach.}
	\label{fig:NSInv}
\end{figure}
 In particular, the inlet velocity field is assumed unknown and will be inferred by assimilating sparse velocity observation data as shown in Fig.~\ref{fig:NSInv}b. The true inlet has a parabolic profile as shown in Fig.~\ref{fig:NSInv}e. The functional form of the profile is not predefined in solving the inverse problem. Namely, the dimension of the inversion is equal to the degrees of free of the inlet, which is more than 20. By assimilating velocity observation data at sparse locations, our proposed method can accurately infer the unknown inlet velocity profile and also recover the entire velocity and pressure fields very well. However, it is observed that inferred inlet from the penalty-based data assimilation approach is not quite accurate, which notably deviates from the ground truth. Despite using same penalty coefficient as the previous cases, the inference performance significantly deteriorates. The proposed way of assimilating data strictly can avoid hyperparameter tuning and have better robustness. 


\section{Conclusion}
\label{sec:Conclusion}
In this paper, a novel discrete PINN framework is proposed for solving both forward and inverse problems governed by PDEs in a unified manner. Built upon the combination of graph convolutional networks (GCNs) and Galerkin variational formulation of physics-informed loss functions, the proposed PINN can naturally handle irregular domains with unstructured meshes, where the training is performed in an efficient way due to the reduced search space by polynomials. Thanks to the hard enforcement of boundary conditions and sparse observation data, the proposed method does not require tuning penalty parameters and has better robustness. The numerical results from several forward and inverse problems governed by linear and nonlinear PDEs have shown the effectiveness of the proposed method. Furthermore, the authors believe this work contributes to facilitating the healthy combination of scientific deep learning and classic numerical techniques rather than isolating them against each other.


\section*{Compliance with Ethical Standards}
Conflict of Interest: The authors declare that they have no conflict of interest.

\section*{Acknowledgment}
The authors would like to acknowledge the funds from National Science Foundation under award numbers CMMI-1934300 and OAC-2047127 (JXW, HG), the Air Force Office of Scientific Research (AFOSR) under award number FA9550-20-1-0236 (MZ), and startup funds from the College of Engineering at University of Notre Dame in supporting this study.


\appendix


\end{document}

%% file: preamble.macros.tex
\usepackage{bm} 
\usepackage{amsfonts} 

\newcommand{\Ecal}{\ensuremath{\mathcal{E}}}
\newcommand{\Fcal}{\ensuremath{\mathcal{F}}}
\newcommand{\Gcal}{\ensuremath{\mathcal{G}}}
\newcommand{\Hcal}{\ensuremath{\mathcal{H}}}

\newcommand{\Lcal}{\ensuremath{\mathcal{L}}}

\newcommand{\Ncal}{\ensuremath{\mathcal{N}}}

\newcommand{\Pcal}{\ensuremath{\mathcal{P}}}

\newcommand{\Vcal}{\ensuremath{\mathcal{V}}}

\newcommand{\Rbb}{\ensuremath{\mathbb{R} }}

\newcommand\Abm{{\ensuremath{\bm{A}}}}

\newcommand\Dbm{{\ensuremath{\bm{D}}}}

\newcommand\Fbm{{\ensuremath{\bm{F}}}}

\newcommand\Ibm{{\ensuremath{\bm{I}}}}

\newcommand\Lbm{{\ensuremath{\bm{L}}}}

\newcommand\Rbm{{\ensuremath{\bm{R}}}}

\newcommand\Ubm{{\ensuremath{\bm{U}}}}

\newcommand\Wbm{{\ensuremath{\bm{W}}}}
\newcommand\Xbm{{\ensuremath{\bm{X}}}}

\newcommand\Zbm{{\ensuremath{\bm{Z}}}}

\newcommand\bbm{{\ensuremath{\bm{b}}}}

\newcommand\fbm{{\ensuremath{\bm{f}}}}

\newcommand\chibold{{\ensuremath{\boldsymbol{\chi}}}}

\newcommand\mubold{{\ensuremath{\boldsymbol{\mu}}}}

\newcommand\Phibold{{\ensuremath{\boldsymbol{\Phi}}}}

\newcommand\Thetabold{{\ensuremath{\boldsymbol{\Theta}}}}


%% file: preamble.tikz.tex
\usepackage{tikz}
\usepackage{pgfplots}
\usepackage{pgfplotstable, filecontents, booktabs}
\pgfplotsset{compat=1.9}

\usetikzlibrary{pgfplots.groupplots}
\usepgfplotslibrary{fillbetween}
\usetikzlibrary{calc,fit,matrix,arrows,automata,positioning,shapes}
\usetikzlibrary{arrows.meta}

\pgfplotsset{select coords between index/.style 2 args={
    x filter/.code={
        \ifnum\coordindex<#1\fi
        \ifnum\coordindex>#2\fi
    }
}}

\tikzset{
 invisible/.style={opacity=0},
 visible on/.style={alt={#1{}{invisible}}},
 alt/.code args={<#1>#2#3}{%
   \alt<#1>{\pgfkeysalso{#2}}{\pgfkeysalso{#3}}
 },
}
